\documentclass{aa}  
\usepackage{longtable}
\usepackage{caption}

\usepackage{graphicx}
\usepackage{txfonts}
\usepackage[fleqn]{amsmath}	
\usepackage{amssymb}	
\usepackage{mathtools}
\usepackage[breaklinks=true]{hyperref}
\hypersetup{
  colorlinks=true,   
  citecolor=blue,    
  urlcolor=blue,     
  linkcolor=blue,     
}
\bibpunct{(}{)}{;}{a}{}{,} 
\usepackage{lscape}

\newcommand{\ks}{K_{\rm S}}

\DeclareMathOperator{\DD}{\mathcal{D}}
\DeclareMathOperator{\B}{\mathcal{B}}

\newcommand\RR{\mathbb{R}}
\newcommand\id{\mathbb{1}}

\DeclareRobustCommand{\ion}[2]{%
\relax\ifmmode
\ifx\testbx\f@series
{\mathbf{#1\,\mathsc{#2}}}\else
{\mathrm{#1\,\mathsc{#2}}}\fi
\else\textup{#1\,{\mdseries\textsc{#2}}}%
\fi}

\begin{document} 

   \title{Non-parametric analysis of the Hubble Diagram with Neural Networks}

   \author{Lorenzo Giambagli\inst{1,2,3}
   \thanks{\email{lorenzo.giambagli@unifi.it}},
    Duccio Fanelli\inst{1,3}
    Guido~Risaliti\inst{1,4},
    Matilde~Signorini\inst{1,4}
          }
          
\institute{
$^{1}$Dipartimento di Fisica e Astronomia, Universit\`a di Firenze, via G. Sansone 1, 50019 Sesto Fiorentino, Firenze, Italy\\
$^{2}$naXys - Namur Center for Complex Systems, University of Namur, rue Graf\'e 2, 5000 Namur, Belgium\\
$^{3}$INFN and CSDC, Via Sansone 1, 50019 Sesto Fiorentino, Firenze, Italy\\
$^{4}$INAF -- Osservatorio Astrofisico di Arcetri, Largo Enrico Fermi 5, I-50125 Firenze, Italy\\
}

\titlerunning{Neural Network Regression Analysis of the Hubble Diagram}
\authorrunning{L. Giambagli et al.}
\date{\today}

 
   \abstract{The recent extension of the Hubble diagram of Supernovae and quasars to redshifts much higher than 1 
prompted a revived interest in non-parametric approaches to test cosmological models and to measure the expansion rate of the Universe.
In particular, 
it is of great interest to infer model-independent constraints on the possible evolution of the dark energy component. Here we present a new method, based on a Neural Network Regression, to analyze the Hubble Diagram in a completely non-parametric, model-independent fashion. We first validate the method through simulated samples with the same redshift distribution as the real ones, and discuss the limitations related to the “inversion problem” for the distance-redshift relation. We then apply this new technique to the analysis of the Hubble diagram of Supernovae and quasars. We confirm that the data up to $ z\sim1-1.5 $ are in agreement with a flat $\Lambda$CDM model with $\Omega_M\sim$0.3, while $\sim5$-sigma deviations emerge at higher redshifts. A flat $\Lambda$CDM model would still be compatible with the data with $\Omega_M>$ 0.4. Allowing for a generic evolution of the dark energy component, we find solutions suggesting an increasing value of $\Omega_M$ with the redshift, as predicted by interacting dark sector models.
}
   

   \maketitle
%

\section{Introduction}
The Hubble diagram (i.e. the distance-redshift relation) describes the expansion of the Universe with time, and is one of the fundamental tools of observational cosmology. The “kinematic” information encoded in this diagram include the Hubble parameter $H_0$ (from the first-order derivative at redshift $z=0$) and the acceleration parameter (from the second-order derivative). When a dynamical model is adopted, its physical parameters can be derived from the fit of the Hubble diagram. Typical examples are the estimate of the matter density at $z=0$, $\Omega_M$, within a flat $\Lambda$CDM model, or the evaluation of $\Omega_M$ and $\Omega_\Lambda$ within a non-flat $\Lambda$CDM model. Moreover, the physical meaning of the relevant parameters is to some extent reflecting the chosen model. Likewise, the obtained numerical estimates are also model-dependent: assume for example data to follow a $\Lambda$CDM model, with prescribed $\Omega_M$ and non-zero curvature. Then, it is easy to demonstrate through numerical simulations that, if a flat $\Lambda$CDM is adopted, the best fit value of $\Omega_M$ will be different from the correct (simulated) one. 

In the past few years, possible new physics beyond the flat $\Lambda$CDM model has been suggested by several observational results, such as the mismatch between the direct measurements of $H_0$ in the local Universe (\citealt{riess2019, Wong19}) and the extrapolations based on the Cosmic Microwave Background (CMB), the comparison between the high- and low- multipole spectra of the CMB (\citealt{DiValentino:2020hov}), and the tension between the power spectrum of density perturbations measured on different scales (\citealt{PhysRevLett.111.161301_sigma8, PhysRevD.91.103508_sigma8, PhysRevD.96.023532_sigma8, S8_Heymans_21, Nunes_21_S8}).  Recently, a significant deviation from the flat $\Lambda$CDM model has been observed in the Hubble diagram at high redshift, populated with quasars and gamma-ray bursts (GRB): while no significant tension is found at  $z<1.5$ with either supernovae, quasars, or GRB, the data at $z>1.5$ suggest a slower expansion of the Universe than predicted by the flat $\Lambda$CDM model (\citealt{rl19, 2020samplelusso2020}).
These results make it particularly important to analyze the Hubble diagram in a model-independent, non-parametric way, in order to obtain an “absolute scale” for the comparison with specific models, and to infer the global, “cosmographic” properties of the expansion which, in turn, could suggest the optimal class of models to fit to the data. 

Cosmographic expansions (\citealt{Aviles2014PrecisionCW_cosmography, 10.1093/mnras/staa871_cosmography, bargiacchi21}) represent a viable approach to pursue this goal. The method is based on a standard fitting procedure and assumes that observational data can be interpolated by an appropriate series of functions, truncated to include a limited number of terms (hence of free parameters). While this is not dependent on a specific physical model, it still relies on the flexibility of the chosen functions to reproduce the shape of the observational Hubble diagram. 

An example of a robust, well checked, non-parametric approach is that based on Gaussian Process regression (\cite{PhysRevD.82.103502_GP}, \cite{Seikel2012ReconstructionOD_GP}, \cite{PhysRevD.85.123530_GP}), which has been used  to test the hypothesis of a constant density of the dark energy term (i.e. the cosmological constant $\Lambda$).

Starting from these premises, we here propose, 
and consequently apply, a novel analysis framework for the Hubble diagram, based on Neural Network Regression. 

We will first describe the method, and check its reliability with simulated data sets. Then we will apply it to a Hubble diagram at high redshift, showing a high-redshift inconsistency with the $\Lambda$CDM model.  Finally we will speculate on the class of models that could fix the discrepancy.


\section{The cosmological background}
\label{obs_data_red}

In a Friedmann-Robertson-Walker Universe, the \textit{luminosity distance} of an astrophysical source is related to the redshift through the equation:
\begin{equation}\label{d}
    d_L=\frac{c\left(1+z\right)}{H_{0}\sqrt{-\Omega_K}}\sin{\left(\sqrt{-\Omega_K}\int_{0}^{z}dz^\prime\frac{H_0}{H\left(z^\prime\right)}\right)}
\end{equation}
where $H(z)$ is the Hubble function and $\Omega_K$ stands for the curvature parameter, defined as $\Omega_K=1-\sum_i\Omega_i$, with $\Omega_i$ representing the density of the constituents of the Universe, normalized to the closure density. 
In the simplest form, assuming a flat Universe, a constant total content of matter in the Universe, a cosmological constant, and considering the redshift range where standard candles are observed (i.e. z $<$ 7, where the contribution of the radiation and neutrino terms is negligible), $H\left(z\right)=H_0\sqrt{\Omega_M\left(1+z\right)^3+1-\Omega_M}$.
However, a wide range of different physical and cosmological models have been considered, including a non-zero curvature, an evolving dark energy density, and/or interactions between dark energy and dark matter. 
In this work, we want to analyze a subset of these models, represented by the equation:
\begin{equation}\label{H}
    H\left(z\right)=H_0\ \sqrt{\Omega_M\left(1+z\right)^3+\left(1-\Omega_M\right)e^{3\int_{0}^{z}{\frac{1+w\left(z^\prime\right)}{1+z^\prime}dz^\prime}}}
\end{equation}
where $w(z)$  is a generic redshift evolution of the dark energy component density. 
Our main goal is to test the consistency of the flat $\Lambda$CDM hypothesis (which amounts to setting $ w=-1 $, in the previous equation) with the present Hubble diagram of supernovae and quasars, and draw comparison with other possible functional forms for $w(z)$, as proposed in the literature. To this aim, we will carry out a non-parametric fit, via a suitably designed Neural Network. This latter enables us to reach conclusions on the predicted profile of $w(z)$ without resting on any a-priori assumption.


One key problem in any non-parametric reconstruction attempt is the so-called "inversion problem":  it is easy to demonstrate that the inversion of Equation (\ref{H}), which involves the first and second derivatives of $H(z)$ (see e.g. \citealt{Seikel2012ReconstructionOD_GP}), is inherently unstable, due to strong dependence on the $\Omega_M$ and $H_0$ parameters (in particular, a change of the quantity $H_0^2\Omega_M$ by as little as 0.1\% can alter the predicted   value of $w(z)$  by orders of magnitude, and/or flip its sign). 
As a consequence, 
constraints on $w(z)$  at very low redshift can be obtained, but the uncertainties become very large already at $ z\sim0.5$. This makes it hard to reach conclusive evidences about the supposed consistency of available data  with the reference scenario with $w=-1$. In principle, better data could help to reduce the uncertainties. While we will discuss this issue in more detail in a dedicated paper, here we just mention the relevant point for the present work: it is not possible to obtain significant information on $w(z)$  from the Hubble diagram without (a) assuming some analytic form of the function and/or (b) having a combined estimate of $\Omega_M$ and $H_0$ with a much higher precision than available today and in the foreseeable future. There are only two possible direct ways to overcome this limitation: either we restrict our analysis to very narrow ranges of the parameters, or we constraint the shape of the function $ w(z) $. Since neither of these approaches is satisfactory (and both of them have been already explored in the literature), we chose a different strategy. We do not attempt to carry out a full inversion of Eq. (\ref{H}). On the contrary, we overcome the aforementioned numerical problems by aiming at estimating the quantity:
\begin{equation}\label{I}
	I\left(z\right)=\int_{0}^{z}\frac{w\left(z^\prime\right)+1}{1+z^\prime}dz^\prime
\end{equation}
which can be determined from the observational data by solely invoking the first derivative of $ H(z) $. We notice that within the $\Lambda$CDM model, $ w=-1 $ implies $ I(z)=0 $.
As an obvious limitation, we will just recover the integral of the physical quantity of interest, the function $ w(z) $: the degeneracy on $w(z)$  implies that different forms of $w(z)$  lead to indistinguishable shapes of $ I(z) $. Nonetheless, we can achieve some remarkable results. First, we can compare the results on $I(z)$  with the prediction of the flat $\Lambda$CDM model: an inconsistency in this check would be a powerful and general proof of a tension between the model and the data (note that the opposite is not true: an agreement based on the analysis of $I(z)$  does not necessarily imply an invalidation of the $\Lambda$CDM model).  More in general, we can explore the family of $w(z)$  functions leading to the observation-based reconstruction of $ I(z) $, to determine which class of physical models can reproduce the observed Hubble diagram. \\
\noindent
\section{Regression via Deep Neural Networks (NN)} 

For our purposes we have chosen to deal with a fully connected feedforward architecture, as illustrated in annexed Supplementary Information (SI).
Function \eqref{I} is hence approximated by a suitable NN, denoted with $ I_{NN} $, to be determined via an apposite optimization procedure,  hereafter outlined. After a few manipulations, as detailed in the SI, the dataset takes the form $ \mathcal{D} = \{(z^{(i)},y^{(i)},\Delta y^{(i)})\}$ with $ i \in 1 \dots |\mathcal{D}| $ where $ y^{(i)} $ is connected to the modulus of luminosity distance $ d_L^{(i)} $ and $ \Delta y^{(i)} $ stands for the associated empirical error.  The predictions $ y_{\text{pred}}^{(i)} $ and the supplied input $ y^{(i)}$ are linked via:

\begin{equation}\label{ypred}
	y_{\text{pred}}^{(i)} = \int_{0}^{z^{(i)}}dz'\left[\Omega_M\left(1+z'\right)^3+\left(1-\Omega_M\right)e^{I_{NN}(z')}\right]^{-\frac{1}{2}}
\end{equation}
Notice that the prediction is a functional of $ I_\text{NN} $, the neural network approximation that constitutes the target of the analysis. To carry out the optimization we introduce the loss function 
$
	L(I_{NN}, \DD) = \sum_{i =1}^{|\DD|}\left(\frac{y^{(i)}-y_{\text{pred}}^{(i)}}{\Delta y^{(i)}}\right)^2 
$. 
The weights of the network which ultimately defines  $I_\text{NN}$  are tuned so as to minimize the above loss function, via  conventional stochastic gradient descent methods. The \textit{hyper-parameters} have been optimized with mock data samples, as illustrated in the SI. To quantify the statistical errors $\Delta y_{\text{pred}}$ (associated to the predictions) and $ \Delta I_\text{NN} $ (referred to the approximating neural network) we implemented a \textit{bootstrap} procedure, further detailed in the SI. The  code is freely available at \url{https://github.com/Jamba15/Cosmological-Regression-with-NN.git}.\\

The regression scheme introduced above was challenged against a selection of mock data samples. In carrying out the test we considered:

{\bf (A)} A sample of 4,000 sources with no dispersion, with a flat distribution in $ \log(z) $ between $ z=0.01 $ and $ z=6 $, and following a flat $\Lambda$CDM model with $\Omega_M=0.3$ and $h=H_0/(100 km/s/Mpc)=0.7$. This sample (as well as the next in the list) represents a highly idealized, hence non realistic setting. It is solely used as a reference benchmark model, for preliminary consistency checks.\\
{\bf (B)} The same as above, but the model used is a Chevallier-Polarski-Linder (CPL) parametrization, (which assumes a Dark Energy equation of state that varies with the redshift as $w(z) = w_0 + w_a \frac{z}{1+z}$ \citep{CHEVALLIER_2001}), with $ w_0=-1.5 $ and $ w_a=0.5 $. \\
{\bf (C)} A sample with the same size, redshift distribution and dispersion as the Pantheon supernovae Ia sample  \citep{scolnic2018}, assuming a flat $\Lambda$CDM model with $\Omega_M =0.3$.\\
{\bf (D)} A Pantheon-like sample, as above, assuming a CPL model with $ w_0=-1.5 $ and $ w_a=0.5 $.\\
{\bf (E)} A sample with the same size and redshift distribution as the combined Pantheon \citep{scolnic2018} and quasar  \citep{2020samplelusso2020} samples. The quasar sample consists of 2,420 sources with redshift in the $ z = 0.5-7.5 $ range. We assume the same dispersion as in the real sample and  a flat $\Lambda$CDM model with $\Omega_M$ =0.3.\\
{\bf (F)} The same as above, assuming a CPL model with $ w_0=-1.5 $ and $ w_a=0.5 $.

More specifically, we generated synthetic data following the different recipes evoked above. The regression scheme, as implemented via the neural network, enables us to solve an inverse problem, from  data back to the underlying physical model. The correspondence between postulated and reconstructed physical instances, readily translates in a reliable metric to gauge the performance of the proposed procedure, in a fully controllable environment and prior application to the experimental dataset. 

The analysis of settings A and B is discussed in the SI, and confirms that our NN method can consistently recover the ``true" model and parameters with simulated data of (unrealistic) high quality. 

\begin{figure}[h]
	\centering
	\includegraphics[scale=.58]{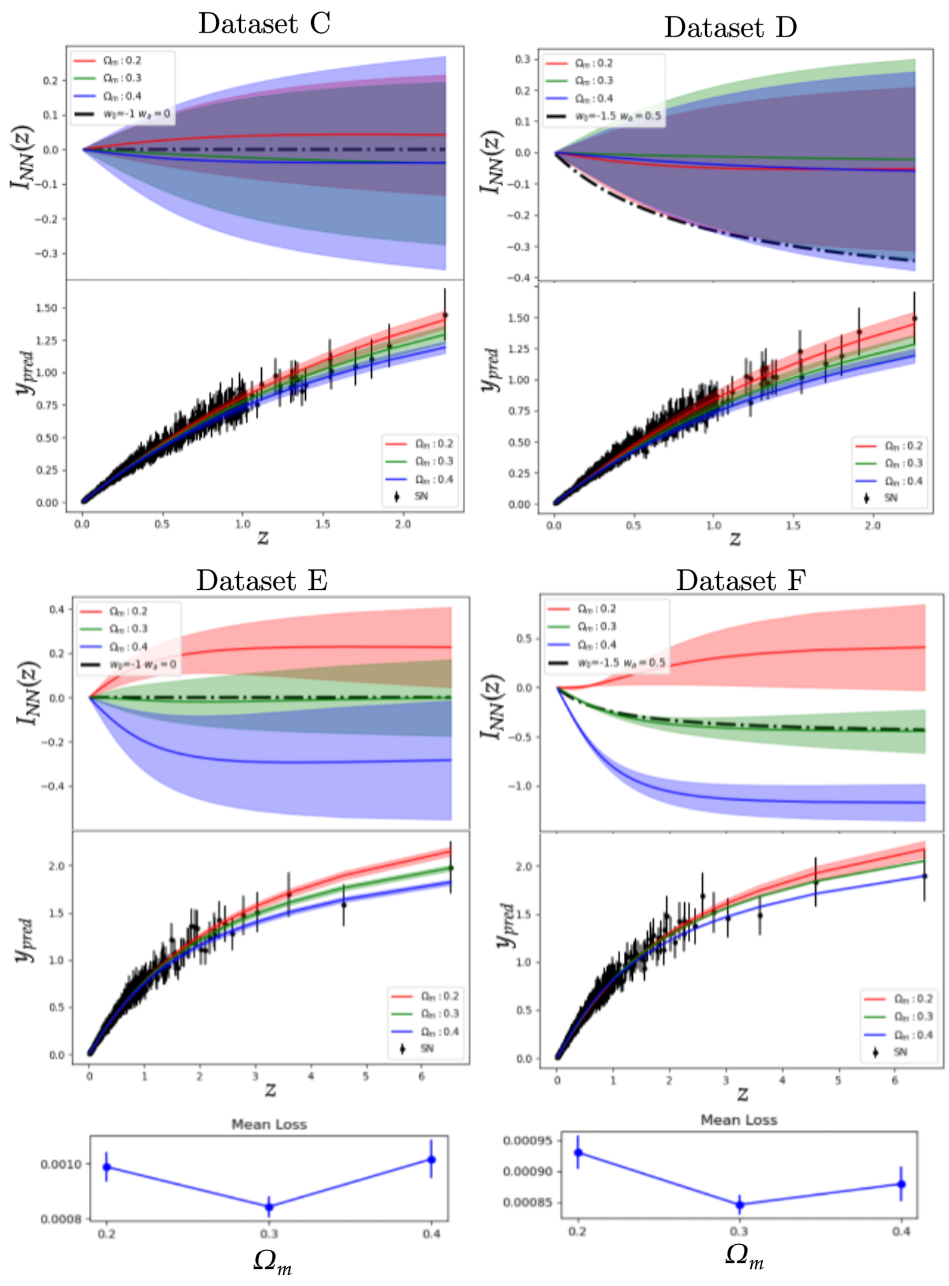}
	\caption{$I(z)$ Results of the NN analysis of the Hubble diagram of simulated data. Top left: Dataset C, with the same redshift distribution and dispersion as the Pantheon supernovae sample. Bottom left: Dataset E, where combined
	Pantheon and quasars are considered. In this case the NN is able to identify the model assumed for data generation (the green shadowed region contains the exact profile for $I_{NN}(z)$, depicted with a dashed line). The corresponding loss function is also shown and displays a minimum at the correct value of $\Omega_M$.  Top right: a Pantheon-like sample is assumed, for a CPL generative model (dataset E). The NN is unable to distinguish between different scenarios ($\Lambda$CDM vs, CPL). Bottom right: CPL model with the inclusion of quasars. The degeneracy is resolved and the NN can correctly identify the underlying model (see dashed line). The loss shows a minimum for the correct value of 
	$\Omega_M$, which yields the green shadowed solution for 
	$I_{NN}(z)$ vs. $z$.}
	\label{f:DatasetCE}
\end{figure}

The outcome of the analysis for respectively settings C (top left), D (top right), E (bottom left) and F (bottom right) is displayed in Figure \ref{f:DatasetCE}. Both $I_{NN}(z)$ (the neural network approximation for $I(z)$) and $y_{pred}(z)$ are represented as function of the redshift $z$. For settings E and F, the associated mean loss is also plotted against the parameter $\Omega_M$, which can be freely modulated to explore different scenarios. Working with a dataset of type C cannot yield definite conclusions: indeed the NN is unable to recover the correct value of $\Omega_M$, as
different $\Lambda$CDM models ($I_{NN}(z) \simeq 0$, within the explored range) provide an equally accurate interpolation of the (simulated) data within statistical errors. The above degeneracy is however removed when extending the examined sample so as to include quasars, see bottom-left panel of Figure \ref{f:DatasetCE} which refers to dataset E. In this case, the minimum displayed by the loss function points to $\Omega_M=0.3$, the value assumed in the simulations, and the corresponding function $I_{NN}(z)$ is approximately equal to zero (green shadowed domain) within errors, and at variance with what it is found by employing the other chosen values of $\Omega_M$. Datasets D and F (rightmost panels in Figure \ref{f:DatasetCE})
returns similar conclusions when operating with data generated according to a CPL prescription. Working with supernovae (over a limited range in $z$) does not allow to distinguish between  $\Lambda$CDM and CPL model, while the underlying model, assumed for data generation, is correctly singled out when quasars are accounted for (green shadowed region that encloses the dashed line, that represents the exact profile), i.e. when extending the dataset to higher redshifts. 
Overall, working on synthetic data suggest that (a) the regression method is reliable, (b) with the current Hubble diagram of supernovae it is not possible to test the $\Lambda$CDM  model against possible extension such as the CPL model with ``phantom like" dark energy. Such a degeneracy is removed with a combined supernovae+quasar sample extending up to z$\sim$7.

\begin{figure}
	\centering
	\includegraphics[scale=0.45]{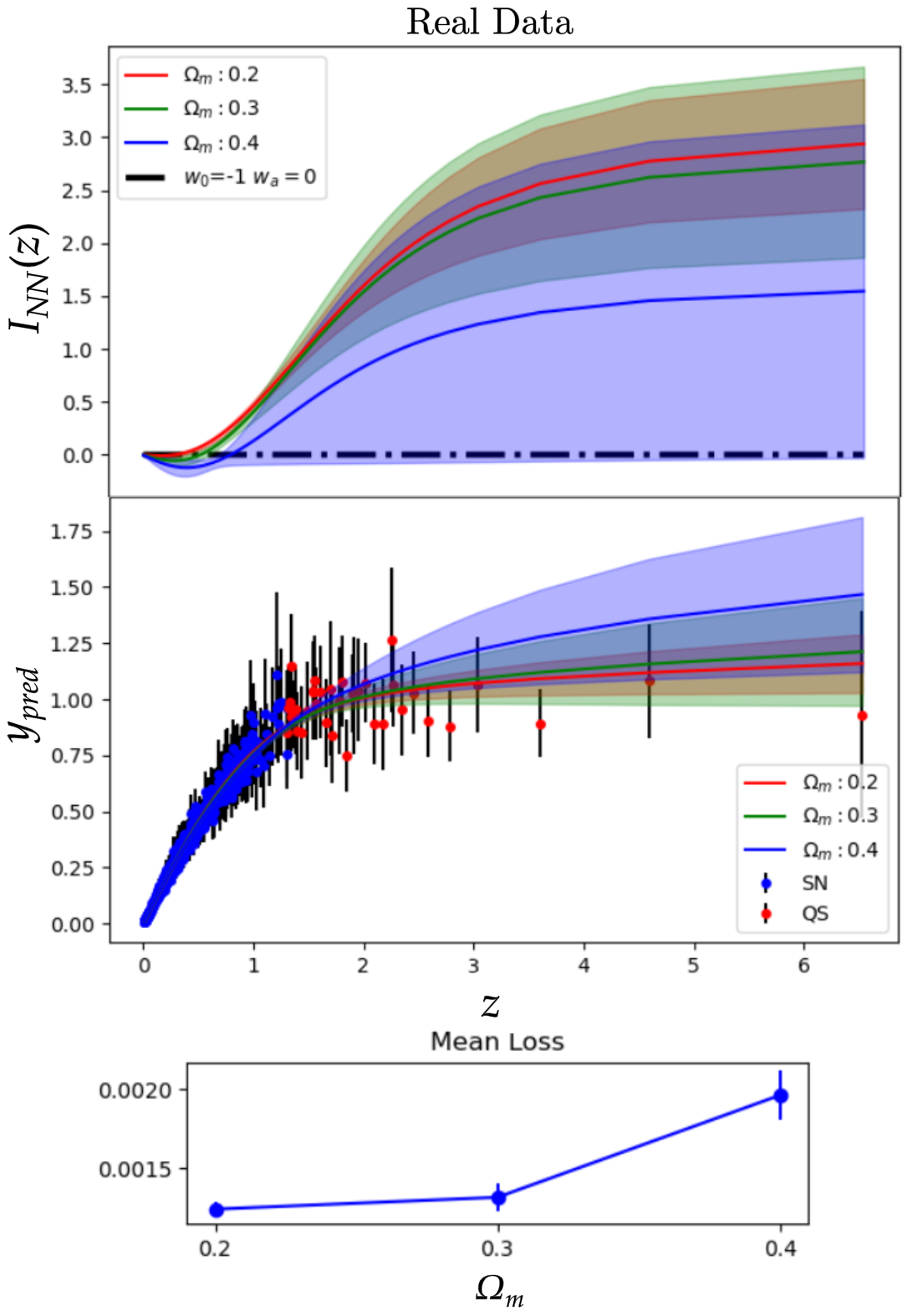}
	\caption { Results of the NN analysis of the Hubble diagram of supernovae (blue points in the middle panel) and quasars (red points). Top panel: estimated values of $I(z)$  for different values of $\Omega_M$. Central panel: Hubble diagram with the reconstructed best fit function obtained from the NN analysis. Bottom panel: $Loss$ values for different values of $\Omega_M$.  Notice that the solution visually closer (accounting for statistical errors) to the reference $\Lambda$CDM profile yields significantly larger value of the loss, and as such should be disregarded. The $Loss$ is indeed nearly flat for $\Omega_M<0.3$. }
	\label{f:realdata} 
\end{figure}

Motivated by this, we applied the NN to the experimental dataset (Pantheon+ quasaris sample) and obtained the results shown in Figure \ref{f:realdata}. The shape of $I(z)$  is clearly non consistent with the flat $\Lambda$CDM model ($I(z)\equiv0$). This is the main result of our work, and has been obtained without assuming any a priori knowledge on the function $I(z)$. 


As a next step in the story, we introduce a dedicated  indicator to quantitatively measure the compatibility of the 
examined data with the reference $\Lambda \text{CDM}$ model. Imagine to naively access the distance of the fitted profile $y_{pred}$ to the reference $y_{\Lambda \text{CDM}}$  ($I=0$) curve and divide it with the error associated to the fitted function  $\Delta y_{pred}$. Assume that the computed ratio (averaged over $z$) is smaller than unit. Then, the distance between  $y_{pred}$ and $y_{\Lambda \text{CDM}}$ is 
eclipsed by statistical uncertainty and thus $\Lambda \text{CDM}$ cannot be ruled out as a candidate explanatory model. The above procedure can be cast on solid grounds (see SI), yielding a scalar indicator that fulfills the purpose of quantifying the sought distance, normalized to the associated error. This is denoted by $ \Delta_{\Lambda \text{CDM}} $ and takes the form: 

\begin{equation}\label{indicator_1}
	\Delta_{\Lambda \text{CDM}}(\DD, I_\text{NN})=\dfrac{1}{|\DD|}\sum_{i\in\DD} \dfrac{\delta y_\text{pred}^{\Lambda \text{CDM}}(I_\text{NN}; z^{(i)})}{\Delta y_\text{pred}(I_\text{NN}; z^{(i)})}
\end{equation}
The fitted integral function $ I_\text{NN} $ is deemed compatible with the $\Lambda \text{CDM}$ model, if $\Delta_{\Lambda \text{CDM}}<1 $. When this latter condition holds true,  the predictions deviate from a $\Lambda \text{CDM}$ by an amount that, on average, is smaller than the corresponding prediction error. The indicator in \eqref{indicator_1} has been computed for different mock samples, mimicking  $\Lambda \text{CDM}$, with progressively increasing errors sizes $ \Delta y $. The latter is assumed uniform across data points and varied from zero to 0.15, thus including the value - $ \sim 0.14 $ - that is believed to apply to real data. This information is used as a reference benchmark to interpret the results of the analysis for the Pantheon + quasar experimental dataset. To sum up our conclusions (see SI) the portion of the dataset at small redshift is compatible with a $\Lambda$\text{CDM} model with $ \Omega_m =0.3 $, within statistical errors. Conversely, for $z > 2$ (notably quasars), $ \Delta_{\Lambda \text{CDM}} $, as computed after available experiments, is $ 5\sigma $ away the expected mean value. Hence, accounting for quasars, enables us to conclude that the $\Lambda \text{CDM}$ model is indeed extremely unlikely.

\begin{figure}
	\centering
	\includegraphics[scale=0.6]{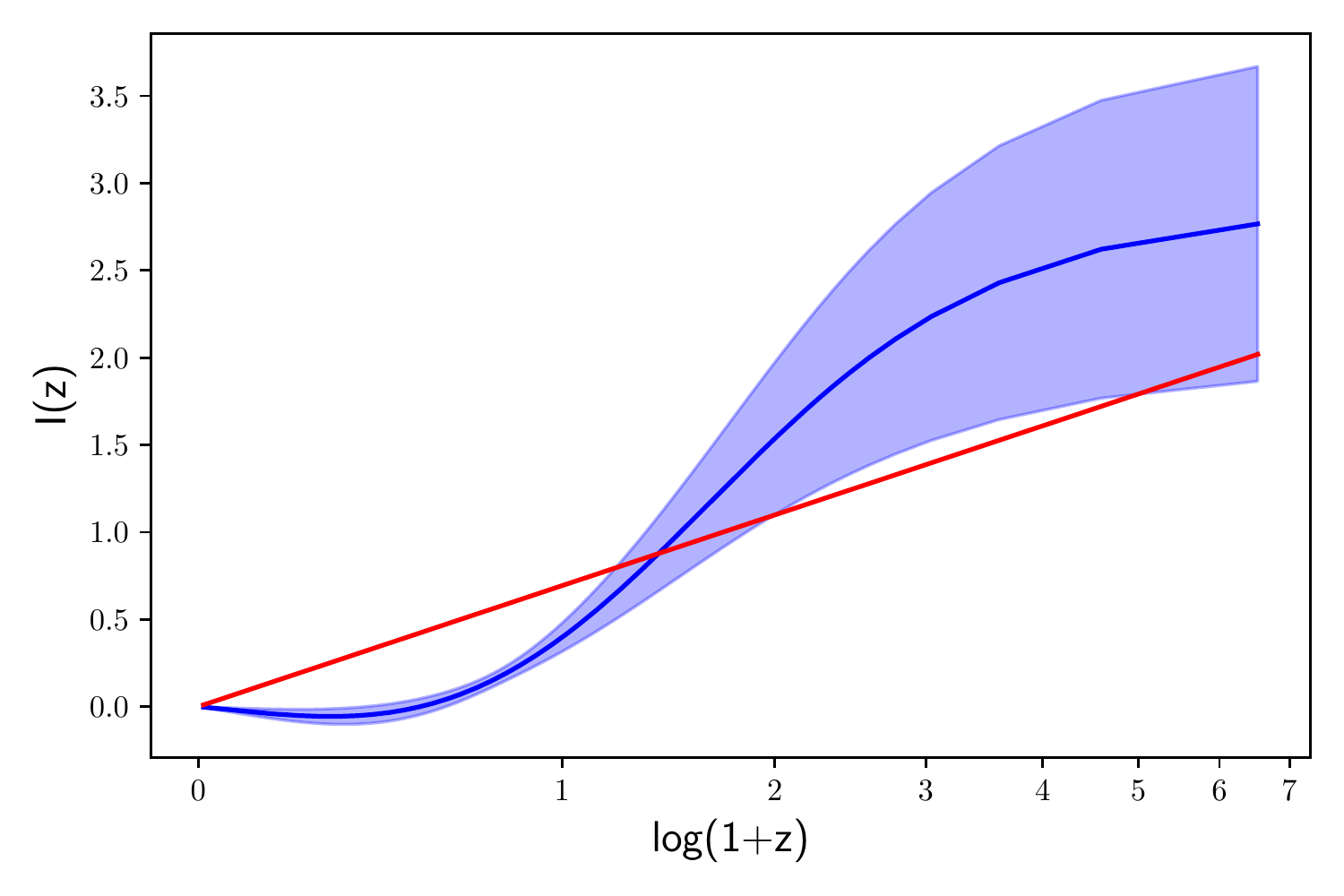} \\ 
	\caption {Best fit $I(z)$ from our NN regression (as in the upper panel of the previous figure) in logarithmic scale, compared with the function $I_{\rm MATTER}(z)$ obtained by assuming $w(z)\equiv0$ in Eq. (\ref{I}). The redshift intervals where the derivative of $I(z)$ is higher than that of $I_{\rm MATTER}(z)$ represent "matter-like" contributions, while intervals with a lower derivative refer to energy-like contributions. }
	\label{f:wref} 
\end{figure}

Finally, we comment on the results depicted in Figure \ref{f:wref} where the best fit $I(z)$ for $\Omega_M=0.3$ (the same as in the upper panel of  Figure \ref{f:realdata}) is plotted in logarithmic scale, and compared to $I_{\rm MATTER}(z)=\log(z)$, the function obtained from equation \eqref{I} by assuming $w(z)\equiv0$, i.e. a pure matter contribution. We recall that a cosmological constant, or equivalently a dark energy component with constant energy, implies $w(z)\equiv-1$ and $I(z)\equiv0$. It is therefore tempting to speculate as follows, when qualitatively analyzing the profile of $I(z)$: the redshift intervals with negative derivative represent a dark energy component with density increasing in time (the ``phantom" dark energy scenario); the intervals with positive derivatives, smaller than the constant derivative of $I_{\rm MATTER}(z)$ represent a dark energy component with decreasing density; last, the intervals where the derivative is larger than that displayed by $I_{\rm MATTER}(z)$ are matter terms, with increasing density. 
The prior-free NN solution suggests therefore an ``interacting dark sector" scenario, where a matter component decreases with time, and correspondingly a dark energy component rises. This interpretation is also consistent with the nearly constant $Loss$, for $\Omega_M<0.3$:  choosing values larger than 0.3  worsen the agreement, because this amounts to overestimate the total matter component at $z\sim0$. On the other hand, value smaller than 0.3 can be compensated by the matter component in $I(z)$.

\section{Conclusions} 
Our conclusions are multifolds. We have proposed and rigorously tested a Neural Network (NN) approach to analyse the Hubble diagram. Then, the NN model-independent regression of the combined supernovae and quasars catalogue enables us to unequivocally reveal a strong tension with the "concordance" flat $\Lambda$CDM model. Finally, the analysis carried out with the proposed NN approach suggests  an ``interacting dark sector" scenario, where a dark matter component flows into dark energy, at least down to redshifts $z\sim1.5$. 

\begin{acknowledgements}
We acknowledge financial contribution from the agreement ASI-INAF n.2017-14-H.O. EL acknowledges the support of grant ID: 45780 Fondazione Cassa di Risparmio Firenze. FS acknowledges is financially supported by the National Operative Program (Programma Operativo Nazionale--PON) of the Italian Ministry of University and Research “Research and Innovation 2014--2020”, Project Proposals CIR01\_00010. A sincere acknowledgement goes to Dr. Colasurdo, who first performed the data reduction of the LBT $\ks$ spectra in her master thesis and developed the baseline analysis that we used as a benchmark. We also acknowledge Prof. Trakhtenbrot for kindly sharing the BASS data. 
\end{acknowledgements}

\bibliographystyle{aa} 
\bibliography{bibl}

\begin{thebibliography}{19}
\expandafter\ifx\csname natexlab\endcsname\relax\def\natexlab#1{#1}\fi

\bibitem[{Aviles {et~al.}(2014)Aviles, Bravetti, Capozziello, \&
  Luongo}]{Aviles2014PrecisionCW_cosmography}
Aviles, A., Bravetti, A., Capozziello, S., \& Luongo, O. 2014, Physical Review
  D, 90, 043531

\bibitem[{{Bargiacchi} {et~al.}(2021){Bargiacchi}, {Risaliti, G.}, {Benetti,
  M.}, {Capozziello, S.}, {Lusso, E.}, {Saccardi, A.}, \& {Signorini,
  M.}}]{bargiacchi21}
{Bargiacchi}, G., {Risaliti, G.}, {Benetti, M.}, {et~al.} 2021, A\&A, 649, A65

\bibitem[{Battye {et~al.}(2015)Battye, Charnock, \&
  Moss}]{PhysRevD.91.103508_sigma8}
Battye, R.~A., Charnock, T., \& Moss, A. 2015, Phys. Rev. D, 91, 103508

\bibitem[{Capozziello {et~al.}(2020)Capozziello, D’Agostino, \&
  Luongo}]{10.1093/mnras/staa871_cosmography}
Capozziello, S., D’Agostino, R., \& Luongo, O. 2020, Monthly Notices of the
  Royal Astronomical Society, 494, 2576

\bibitem[{Chevallier \& Polarski(2001)}]{CHEVALLIER_2001}
Chevallier, M. \& Polarski, D. 2001, International Journal of Modern Physics D,
  10, 213–223

\bibitem[{Di~Valentino {et~al.}(2021)Di~Valentino, Melchiorri, \&
  Silk}]{DiValentino:2020hov}
Di~Valentino, E., Melchiorri, A., \& Silk, J. 2021, Astrophys. J. Lett., 908,
  L9

\bibitem[{{Heymans, C.} {et~al.}(2021){Heymans, C.}, {Tr\"oster, T.}, {Asgari,
  M.}, {Blake, C.}, {Hildebrandt, H.}, {Joachimi, B.}, {Kuijken, K.}, {Lin,
  C.}, {S\'anchez, A. G.}, {van den Busch, J. L.}, {Wright, A. H.}, {Amon, A.},
  {Bilicki, M.}, {de Jong, J.}, {Crocce, M.}, {Dvornik, A.}, {Erben, T.},
  {Fortuna, M. C.}, {Getman, F.}, {Giblin, B.}, {Glazebrook, K.}, {Hoekstra,
  H.}, {Joudaki, S.}, {Kannawadi, A.}, {K\"ohlinger, F.}, {Lidman, C.},
  {Miller, L.}, {Napolitano, N. R.}, {Parkinson, D.}, {Schneider, P.}, {Shan,
  H.}, {Valentijn, E. A.}, {Verdoes Kleijn, G.}, \& {Wolf, C.}}]{S8_Heymans_21}
{Heymans, C.}, {Tr\"oster, T.}, {Asgari, M.}, {et~al.} 2021, A\&A, 646, A140

\bibitem[{Holsclaw {et~al.}(2010)Holsclaw, Alam, Sans\'o, Lee, Heitmann, Habib,
  \& Higdon}]{PhysRevD.82.103502_GP}
Holsclaw, T., Alam, U., Sans\'o, B., {et~al.} 2010, Phys. Rev. D, 82, 103502

\bibitem[{Liaw {et~al.}(2018)Liaw, Liang, Nishihara, Moritz, Gonzalez, \&
  Stoica}]{tune}
Liaw, R., Liang, E., Nishihara, R., {et~al.} 2018, arXiv preprint
  arXiv:1807.05118

\bibitem[{Lin \& Ishak(2017)}]{PhysRevD.96.023532_sigma8}
Lin, W. \& Ishak, M. 2017, Phys. Rev. D, 96, 023532

\bibitem[{{Lusso} {et~al.}(2020){Lusso}, {Risaliti}, {Nardini}, {Bargiacchi},
  {Benetti}, {Bisogni}, {Capozziello}, {Civano}, {Eggleston}, {Elvis},
  {Fabbiano}, {Gilli}, {Marconi}, {Paolillo}, {Piedipalumbo}, {Salvestrini},
  {Signorini}, \& {Vignali}}]{2020samplelusso2020}
{Lusso}, E., {Risaliti}, G., {Nardini}, E., {et~al.} 2020, A\&A, 642, A150

\bibitem[{Macaulay {et~al.}(2013)Macaulay, Wehus, \&
  Eriksen}]{PhysRevLett.111.161301_sigma8}
Macaulay, E., Wehus, I.~K., \& Eriksen, H.~K. 2013, Phys. Rev. Lett., 111,
  161301

\bibitem[{Nunes \& Vagnozzi(2021)}]{Nunes_21_S8}
Nunes, R.~C. \& Vagnozzi, S. 2021, Monthly Notices of the Royal Astronomical
  Society, 505, 5427

\bibitem[{{Riess} {et~al.}(2019){Riess}, {Casertano}, {Yuan}, {Macri}, \&
  {Scolnic}}]{riess2019}
{Riess}, A.~G., {Casertano}, S., {Yuan}, W., {Macri}, L.~M., \& {Scolnic}, D.
  2019, \apj, 876, 85

\bibitem[{{Risaliti} \& {Lusso}(2019)}]{rl19}
{Risaliti}, G. \& {Lusso}, E. 2019, Nature Astronomy, 195

\bibitem[{{Scolnic} {et~al.}(2018){Scolnic}, {Jones}, {Rest}, {Pan},
  {Chornock}, {Foley}, {Huber}, {Kessler}, {Narayan}, {Riess}, {Rodney},
  {Berger}, {Brout}, {Challis}, {Drout}, {Finkbeiner}, {Lunnan}, {Kirshner},
  {Sand ers}, {Schlafly}, {Smartt}, {Stubbs}, {Tonry}, {Wood-Vasey}, {Foley},
  {Hand}, {Johnson}, {Burgett}, {Chambers}, {Draper}, {Hodapp}, {Kaiser},
  {Kudritzki}, {Magnier}, {Metcalfe}, {Bresolin}, {Gall}, {Kotak}, {McCrum}, \&
  {Smith}}]{scolnic2018}
{Scolnic}, D.~M., {Jones}, D.~O., {Rest}, A., {et~al.} 2018, \apj, 859, 101

\bibitem[{Seikel {et~al.}(2012)Seikel, Clarkson, \&
  Smith}]{Seikel2012ReconstructionOD_GP}
Seikel, M., Clarkson, C., \& Smith, M. 2012, arXiv: Cosmology and Nongalactic
  Astrophysics

\bibitem[{Shafieloo {et~al.}(2012)Shafieloo, Kim, \&
  Linder}]{PhysRevD.85.123530_GP}
Shafieloo, A., Kim, A.~G., \& Linder, E.~V. 2012, Phys. Rev. D, 85, 123530

\bibitem[{Wong {et~al.}(2019)Wong, Suyu, Chen, Rusu, Millon, Sluse, Bonvin,
  Fassnacht, Taubenberger, Auger, Birrer, Chan, Courbin, Hilbert, Tihhonova,
  Treu, Agnello, Ding, Jee, Komatsu, Shajib, Sonnenfeld, Blandford, Koopmans,
  Marshall, \& Meylan}]{Wong19}
Wong, K.~C., Suyu, S.~H., Chen, G., {et~al.} 2019, Monthly Notices of the Royal
  Astronomical Society, 498, 1420

\end{thebibliography}

\begin{appendix} 

	\section{Data processing}\label{preprocess}
	
	 Data come as the set $ \mathcal{D} = \{(z^{(i)},y^{(i)},\Delta y^{(i)})\}$ with $ i \in 1 \dots |\mathcal{D}| $. Each component $ y^{(i)} $ is linked to $ d_L $, the physical quantity of interest, by $ y^{(i)} = 5\log(d_L^{(i)}/10$pc$) $. The first applied transformation is defined as follows: 
	\begin{equation}\label{transform 1}
		y'^{(i)} = y^{(i)}/5 + 1, \quad \Delta y'^{(i)} = \Delta y^{(i)}
	\end{equation}  
	By doing so data are traced back to the logarithm of the luminosity distance; every entry of the inspected dataset is indeed equal to $ y^{(i)}=\log(d_L^{(i)}) $.\\
	
	Carrying out a first order expansion of equation (1) in the main body of the paper, assuming a flat Universe ($\Omega_k \sim 0$) and inserting the expression of $ H(z) $ as reported in the main text, yields:
	
	\begin{equation}
		d_L=\alpha(z)\int_{0}^{z}dz'\left[\Omega_M\left(1+z'\right)^3+\left(1-\Omega_M\right)e^{I(z')}\right]^{-\frac{1}{2}}
	\end{equation}
	where $ \alpha(z)=\frac{c\left(1+z\right)}{H_{0}} $. Then we proceed by setting:
	\begin{equation}
		y''^{(i)}=y'^{(i)}-\log(\alpha(z^{(i)})), \quad \Delta y''^{(i)} = \Delta y'^{(i)}
	\end{equation}
	It is worth noticing that the relative errors associated with $ c, z $ and $ H_0 $ are negligible. The above relation transforms into:  
	\begin{equation}
		y'''^{(i)}=10^{y''^{(i)}}, \quad \Delta y'''^{(i)} = 10^{y''^{(i)}} \Delta y''^{(i)}
	\end{equation}
	To simplify the notation we drop the apex by setting $ y''' \rightarrow y $ and obtain the sought connection between every $ y^{(i)} $ and the function to be fitted $ I(z) $, namely:
	\begin{equation}\label{y d rel}
		y^{(i)} = \int_{0}^{z^{(i)}}dz'\left[\Omega_M\left(1+z'\right)^3+\left(1-\Omega_M\right)e^{I(z')}\right]^{-\frac{1}{2}}
	\end{equation}

	\section{The employed Neural Network model}\label{NNmodel}
	To approximate the non linear scalar function $ I(z): z\in \RR \mapsto I(z) \in \RR$ we make use of a so called \textit{feedforward} architecture. The information flow from the input neuron, associated to $ z^{(i)} $ to the output neuron where the predicted value of  $ I_\text{NN}(z^{(i)}) $ is displayed.\\
	The transformation from layer $k$ to its adjacent homologue $k+1$, following a feedfoward arrangement, is characterized by two nested operations: (i) a linear map $ W^{(k)} : \RR^{N_{k}} \rightarrow \RR^{N_{k+1}}$ and (ii) a non linear filter $ \sigma^{(k+1)}(\cdot) $ applied to each entry of the obtained vector. Here $ k $ ranges in the interval $ 1\dots \ell $ where $ N_1=1 $ and $ \ell $ is the number of \textit{layers}, i.e. the depth of the NN. We have chosen $ \sigma^{(k)}:=\tanh, \ \forall k<\ell-1$ whereas $ \sigma^{(\ell)}=\id$. 
	
	The activation of every neuron in layer $ k $ can be consequently obtained as:
	\begin{equation*}\label{key}
		\vec{x}^{(k)}=W^{(k-1)}(\dots \sigma(W^{(2)}(\sigma(W^{(1)}z)))\dots)
	\end{equation*}

	Furthermore, we have fixed $ N_k=N_{k+1} \ \forall k\in 2\dots\ell-2 $, meaning that every layer (but the first and the last) has the same size as the others. The size of the so called \textit{hidden layer} $ N_2 $ and the total amount of layers $ \ell $ are, consequently, the only hyper-parameters to be eventually fixed. \\
	 Occasionally a neuron-specific scalar, called \textit{bias}, can be added after application of each linear map $ W^{(k)} $. To allow for the solution $ I_\text{NN}(0)=0$ to be possibly recovered, we have set the bias to zero. 
	 
	 The output $ I_\text{NN}(z) $ hence depends on  $ N=\sum_{k=1}^{\ell-1}N_k\times N_{k+1} $ free scalar parameters (the weights $ W^{(k)}_{i,j}, i\in 1\dots N_{k+1} \ j\in 1\dots N_{k} , \ k \in 1\dots \ell-1$), that constitute the target of the optimization.

	\section{Model Optimization}\label{optimization}

	The optimization herefter described has been carried out by using parallel computing on GPU \citep{tune}.\\
	The minimization of the Loss function as defined in the main text is performed via a variant of the stochastic gradient descent (SGC) method, recalled below.\\
	First, the dataset $ \DD $ is shuffled and divided into smaller subsets $ \B_i $ of size $ |\B_i|=\beta $. These are the \textit{batches}, and meet the following condition: $ \DD = \sqcup_i^{N_b} \B_i $. Obviously the number of batches $ N_b $ is equal to $ \lceil \frac{|\DD|}{\beta} \rceil$. \\
	The gradient with respect to every weight $ W $ entering the definition of the function $ L $ is computed, within each batch, as:
	\begin{equation}\label{key}
		G^{(i)}=\nabla_W L(W,\B_i) = \nabla_W \sum_{j:y^{(j)}\in \B_i }\left(\frac{y^{(j)}-y_{\text{pred}}^{(j)}(z^{(j)};W)}{\Delta y^{(j)}}\right)^2
	\end{equation}
	While $ i $ takes values in the range $ 1\dots N_b $, the weights $ W $ are updated so as to minimize, via a stochastic procedure, the Loss function. This is achieved as follows:
	\begin{equation}\label{key}
		W \leftarrow W - \alpha G^{(i)}
	\end{equation}
	The hyper-parameter $ \alpha $ is called \textit{ learning rate} and drives the amount of stochasticity in the Loss descent process. In the present work a more complex yet conceptually equivalent variant of the SGD called Adam is implemented.\\
	
	A so called \textit{epoch} is completed when all batches have been used. The number of epochs $ N_e $ is another hyper-parameter that has to be fixed a priori, as well as the batch size $ \beta $.
	Usually a high number of epochs (such as 400 or 600, as employed in the present application) is chosen. To avoid overfitting, the \textit{early stop} technique is employed. Such technical aid consists in taking a small subset,  $ \mathcal{V} $, of the dataset ($ \sim15\% $ of $ \DD $) and exclud it from the training process. During training  stages, hence, the employed dataset is $ \DD' = \DD - \mathcal{V} $. While applying SGD to the Loss so as to minimize it, Loss evaluation on dataset $ \mathcal{V} $, $ L(I_\text{NN}, \mathcal{V}) $ is also performed. When the latter function reaches a plateau, the optimization process is stopped. This latter procedure relies on two hyper-parameters: $ \delta $ the absolute variation of $ L $ that can be considered as a real Loss change, and $ p $, the number of consecutive epochs with no recorded variation, before the fitting algorithm can be eventually terminated.\\
	Moreover, one additional hyper-parameter needs to be mentioned: as already explained in the main body of the paper, the prediction $ y_\text{pred} $ involves a numerical integral of the NN approximating function, $ I_\text{NN} $. The integration step $ dz' $ is thus to be set, and was object of a meticulous optimization.\\	
	An hyper-optimization process designed to find the best set of hyper-parameters has been carried out, employing several CPL and $ \Lambda$CDM like models. Such process has led to a set of parameters which have been fixed and left unchanged during the trials. In Table I the chosen hyper-parameters list is provided.
	\begin{table}[h]
		\centering
		\begin{tabular}{|l|c|c|c|c|c|c|c|}
			{$ N_2 $}&{$ \ell $}&{$ N_e $} & {$ \beta $ }& {$ \alpha $} & {$ \delta $} &  {$ p $} &  {$ dz' $} \\ 
\hline
			{$ 20 $}&{$ 5 $}&{$ 600 $}  &  {$ 100 $}  & {$ 10^{-6}  $} &  {$ 10^{-6} $}   &  {$ 35 $}  & {  $ 5\ 10^{-4 } $ }\\
		\end{tabular}
		\caption{Hyper-parameters employed}
	\end{table}\label{hyperparams}

	\section{Simulations results}\label{results}

In the following we will report about the results of the regression model against the simulated settings mentioned, but not displayed, in the main text. 

\begin{figure}[h]
	\centering
	\includegraphics[scale=1.3]{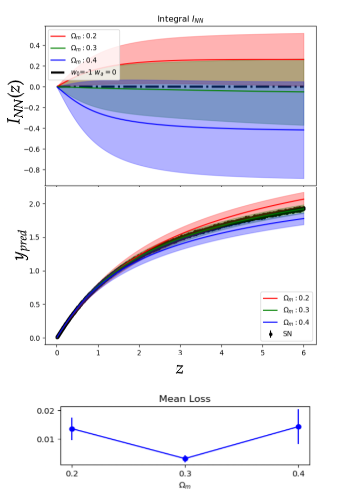}\\ 
	\caption{{ Simulations with a "perfect" sample, dataset A.} { Results of the NN analysis of a simulated sample of 4,000 objects with a log-flat redshift distribution and a negligible dispersion with respect to a flat $\Lambda$CDM model with $\Omega_M$=0.3. Top panel: estimated values of $I(z)$  for different values of $\Omega_M$ (Eq. 3, the "correct" value for the simulated data is I(z)$\equiv$0). Central panel: Hubble diagram with the reconstructed best fit function obtained from the NN analysis. Bottom panel: LOSS values for different values of $\Omega_M$. The minimum is at $\Omega_M$=0.3, i.e. the "true" value. The corresponding $I(z)$  is consistent with zero at all redshifts. These results demonstrate that the NN analysis is able to recover the correct model and the "true" value of $\Omega_M$.}}
	\label{f:DatasetA} 
\end{figure}

\begin{figure}
	\centering
	\includegraphics[scale=1.3]{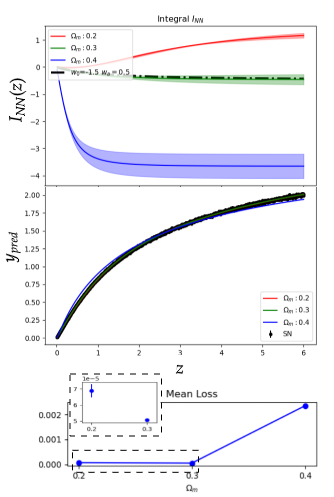}
	\caption{Results for dataset B. The governing  model is a CPL with $w_0=-1.5, w_a=0.5$}
	\label{f:DatasetB}
\end{figure}

	\section{Estimating the errors} \label{exclusion}

To estimate the prediction error $ \Delta y_\text{pred}(z) $ we have employed a  \textit{Bootstrap method}. To this end the fitting procedure is arranged so as to produce $ B $ independent estimators of the quantity $ y_\text{pred} $ and $ I_\text{NN}(z) $, namely $ y_\text{pred}^{[k]} $ and $ I_\text{NN}^{[k]} $ with $ k \in 1\dots B$.
	Each $ y_\text{pred}^{[k]} $ is the result of an optimization process started from a subset $ \DD^{[k]}\subseteq \DD $ obtained from $ \DD $ by uniform sampling with replacement of $ |\DD| $ elements. The prediction errors $ \Delta y_\text{pred} $ and $ \Delta I_\text{NN} $ are then computed by extracting the standard deviation from both sets as:
	\begin{equation}\label{errors}
		\begin{aligned}
		&\Delta y_\text{pred}(z) &= &\sum_{k=1}^{B}\sqrt{\frac{\left(\bar{y}_\text{pred}(z)-y_\text{pred}^{[k]}(z)\right)^2}{B-1}} \\ &\Delta I_\text{NN}(z) &=&\sum_{k=1}^{B}\sqrt{\frac{\left(\bar{I}_\text{NN}(z)-I_\text{NN}^{[k]}(z)\right)^2}{B-1}}
		\end{aligned}
	\end{equation}
	where symbols $ \bar{y}_\text{pred}(z) $ and $ \bar{I}_\text{NN}(z) $ represent the arithmetic mean of the estimates $ y_\text{pred}^{[k]} $ and $ I_\text{NN}^{[k]} $. All across this work, the errors are computed after $ B=80 $ bootstrap samples.

As a next step we shall comment on the derivation of the indicator to gauge the correspondence of the fitted model with a conventional $\Lambda \text{CDM}$ scheme.  We begin by formally expressing $ \delta y_\text{pred}$, the distance of the obtained prediction with respect to the reference $\Lambda \text{CDM}$ model, as
\begin{equation}\label{funcder}
	\begin{aligned}
\delta y_\text{pred}^{\Lambda \text{CDM}}(I_\text{NN}; z) &= \dfrac{\delta y_\text{pred}}{\delta I}\biggr\rvert_{I=\Lambda \text{CDM}}\delta I\\ &= \dfrac{\delta y_\text{pred}}{\delta I}\biggr\rvert_{\Lambda \text{CDM}}(I_\text{NN}-I_{\Lambda \text{CDM}})
	\end{aligned}	 
\end{equation}
where  $ \dfrac{\delta y_\text{pred}}{\delta I}$ stands for the functional derivative and $ I_{\Lambda \text{CDM}}=0 $. The above equation can be further expanded so as to yield:
\begin{equation}
	\dfrac{\delta y_\text{pred}}{\delta I}\biggr\rvert_{I=\Lambda \text{CDM}} = -\dfrac{1}{2}\int_{0}^{z}\alpha(z')^{-\frac{3}{2}}(1-\Omega_m)e^{I(z')}\biggr\rvert_{I=0}
\end{equation}
where $\alpha(z')=\Omega_M\left(1+z'\right)^3+\left(1-\Omega_M\right)e^{I(z')}$. By eventually setting $ \delta I = I_\text{NN} $ one gets therefore: 
\begin{equation}\label{key}
	\delta y_\text{pred}^{\Lambda \text{CDM}}(I_\text{NN}; z) = \dfrac{\Omega_m-1}{2}\int_{0}^{z}\left(\alpha(z')\big\rvert_{I=0}\right)^{-\frac{3}{2}}I_\text{NN}(z')dz'
\end{equation}
 We are finally in a position to introduce the scalar indicator that fulfills the purpose to quantifying the sought distance, normalize to the associated error. This is denoted by $ \Delta_{\Lambda \text{CDM}} $ are takes the form: 

\begin{equation}\label{indicator}
	\Delta_{\Lambda \text{CDM}}(\DD, I_\text{NN})=\dfrac{1}{|\DD|}\sum_{i\in\DD} \dfrac{\delta y_\text{pred}^{\Lambda \text{CDM}}(I_\text{NN}; z^{(i)})}{\Delta y_\text{pred}(I_\text{NN}; z^{(i)})}
\end{equation}
The fitted integral function $ I_\text{NN} $ is deemed compatible with the $\Lambda \text{CDM}$ model, if $\Delta_{\Lambda \text{CDM}}<1 $. When this latter condition holds true,  the predictions deviate from a $\Lambda \text{CDM}$ by an amount that, on average, is smaller than the corresponding prediction error.\\

The indicator in \eqref{indicator} has been computed for different mock samples, mimicking  $\Lambda \text{CDM}$, with progressively increasing errors sizes (assumed uniform across data points), $ \Delta y $  (ranging from zero to 0.15, thus including the value - $ \sim 0.14 $ - that is believed to apply to real data).  

For every choice of the assigned error,  30 mock samples with $ \Omega_m=0.3 $ have been generated and subsequently fitted, assuming different choices of $ \Omega_m $, namely $ \{0.2, 0.3, 0.4 \} $.  For every selected $ \Omega_m $ a bootstrap procedure is implemented (see SI) to estimate $ y_\text{pred}, \Delta y_\text{pred} $ and $ I_\text{NN}, \Delta I_\text{NN} $. The best fit values are selected to be those associated to the smaller mean loss functions (evaluated against the imposed $ \Omega_m $). Following this choice, the mean an the variance of $\Delta_{\Lambda \text{CDM}}$ are computed, from the outcomes of the fits, performed on the corresponding (30) independent realizations. 

\begin{figure}
	\centering
	\includegraphics[scale=0.45]{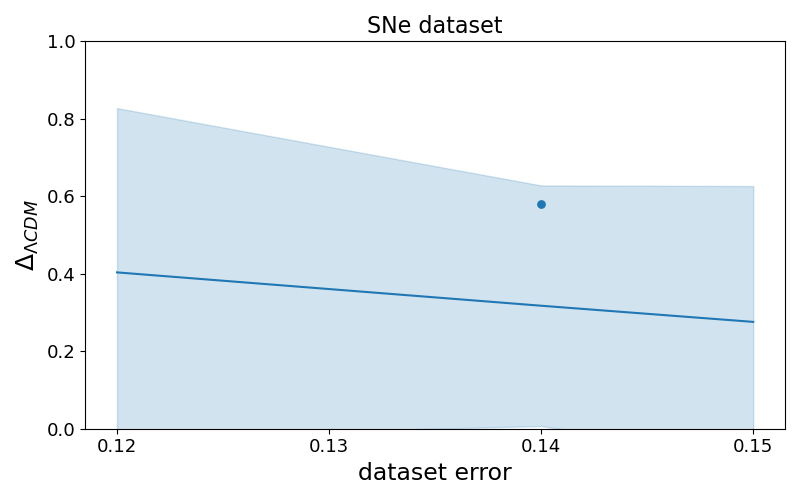}
	\caption{$\Delta_{\Lambda \text{CDM}}$ vs. the imposed error, for the Pantheon dataset (i.e. just supernovae). The symbol stands for to the experimental data, while the solid line and the shadowed regions refer to the corresponding theoretical benchmarks, obtained as described in the text.}
	\label{f:indicatorSN}
\end{figure}

\begin{figure}
	\centering
	\includegraphics[scale=0.38]{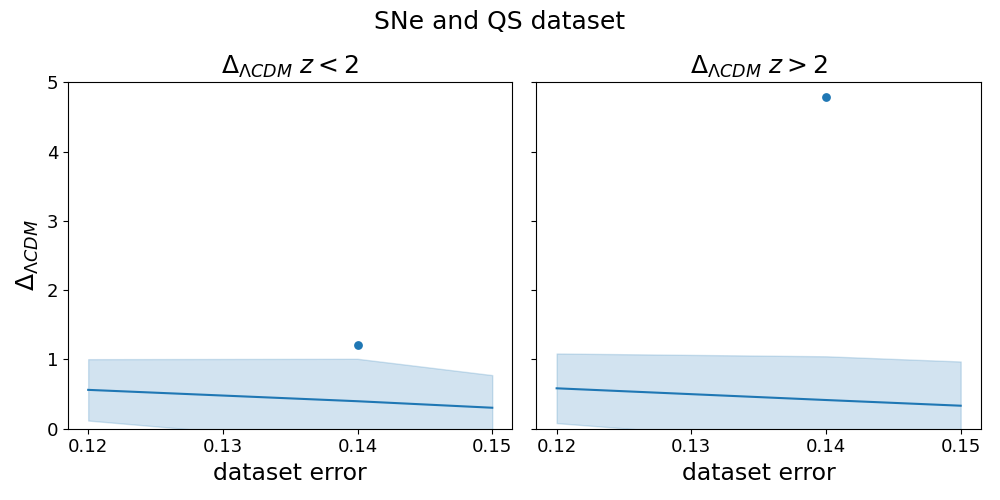}
	\caption{$\Delta_{\Lambda \text{CDM}}$ vs. the imposed error, for the combined supernovae + quasars sample at redshifts $z<2$ (left panel) and $z>2$ (right panel). Symbols refer to the experimental data, while the solid line and the shadowed regions stand for the corresponding theoretical benchmarks, obtained as described in the text.}
	\label{f:indicator}
\end{figure}

\begin{figure}
	\centering
	\includegraphics[scale=0.45]{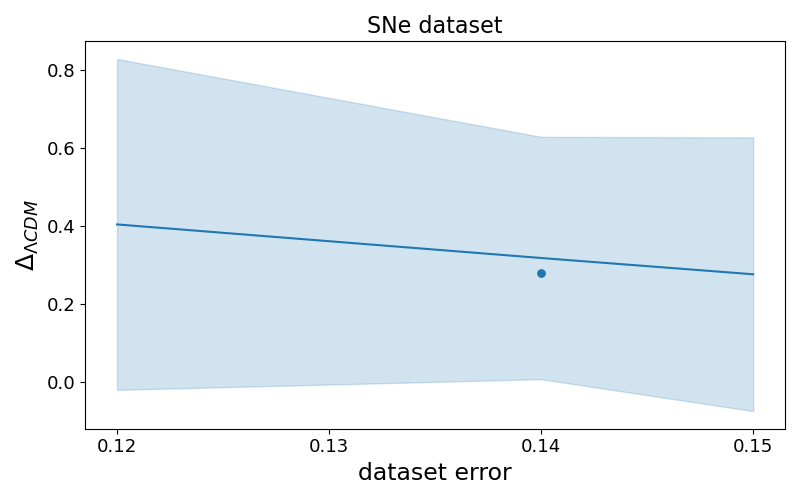}
	\caption{$ \Delta_{\Lambda \text{CDM}} $ vs. the imposed error, for the Pantheon dataset (i.e. just supernovae). In blue the reference mean and variance (represented as a shaded region). The symbol is obtained upon processing the synthetic example generated via the CPL model.}
	\label{f:indicatorSN_CPL}
\end{figure}

\begin{figure}
	\centering
	\includegraphics[scale=0.38]{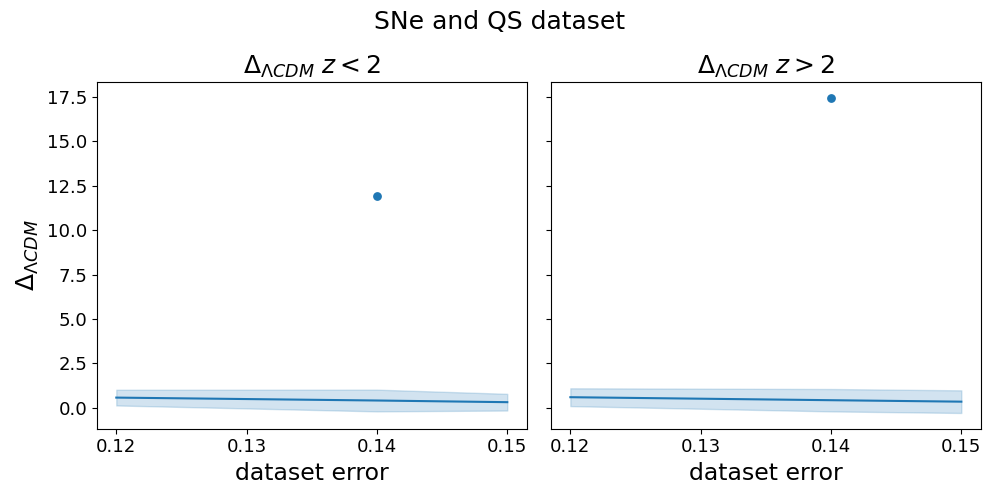}
	\caption{$ \Delta_{\Lambda \text{CDM}} $ vs. the imposed error, for the combined supernovae + quasars sample at redshifts $z<2$ (left panel) and $z>2$ (right panel). In blue the reference mean and variance (represented as a shaded region) obtained with mock $\Lambda \text{CDM}$ samples. The symbols are obtained upon processing the synthetic example generated via the CPL model.}
	\label{f:indicator_CPL}
\end{figure}

In Figures from \ref{f:indicatorSN} to \ref{f:indicator_CPL} the results of the analysis for the different datasets are displayed. The solid line stands for the average estimates, as obtained following the above procedure. The shadowed region is traced after the computed errors, namely, the variance of the indicator across the realizations. 

In Figure \ref{f:indicatorSN} SNe data ($z < 2$) are solely considered for carrying out the regression. The symbol refers to the experimental dataset \citep{2020samplelusso2020} and is set in correspondence of the estimated error (0.14). The displayed point falls within the shadowed domain, thus implying that the examined dataset is compatible with a $\Lambda \text{CDM}$ model. 

In Figure \ref{f:indicator} we analyze the full dataset (Pantheon + quasars). The regression is hence carried out by considering data spanning the whole range in $z$. After the fitting has been performed, data are split into two different regions, respectively at small ($z \leq 2$) or large ($z \geq 2$) redshift. The symbols refers to the experimental dataset and are set in correspondence of the estimated error (0.14). The portion of the dataset at small redshift (mostly populated by Supernovae)  is compatible with a $\Lambda \text{CDM}$ model with $ \Omega_m =0.3 $), within statistical errors (the agreement is even more pronounced if the regression is carried out by solely accounting for Supernovae, see Figure \ref{f:indicatorSN}). Conversely, for $z > 2$, the point computed after available experiments, notably quasars, is at a distance of about $ 5\sigma $ from the expected value of the indicator $ \Delta_{\Lambda \text{CDM}} $. Hence, accounting for quasars enables us to conclude that the $\Lambda \text{CDM}$ model is indeed extremely unlikely.

In Figures \ref{f:indicatorSN_CPL} and \ref{f:indicator_CPL} we repeat the analysis by employing a dataset generated from a CPL model, with an error compatible with that estimated experimentally (equivalent to datasets D and F). The results indicate that  accounting for data at large redshifts is mandatory  to resolve the degeneracy between distinct generative models.

\begin{figure}
    \centering
	\includegraphics[scale=1]{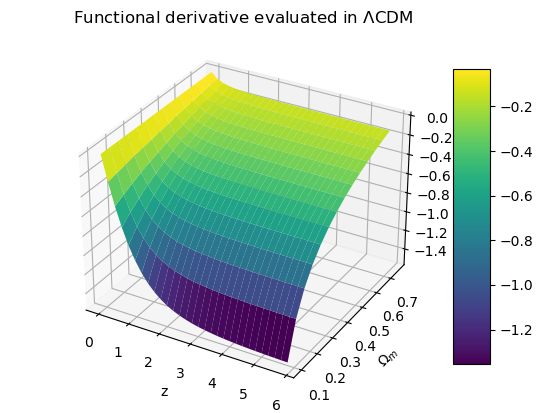}
	\caption{Plot of the functional derivative computed in \eqref{funcder} varying $ \Omega_m $ and $ z $.}
	\label{f:funcder}
\end{figure}

As a final point we elaborate on the reason why different models appear indistinguishable at small 
$z$. Function $ I_\text{NN} $ is the argument of a functional that goes from the space of function $ I $ to the space of the predictions. The way those two spaces communicate (or rather how function $ I $ reverberates on every $ y_\text{pred} $) is a non trivial function of the hyperparameters (as e.g. $ \Omega_m $ and the integration steps) and the domain explored. To clarify this point we plot the functional derivative $ \dfrac{\delta y_\text{pred}}{\delta I}$ (evaluated at $\Lambda \text{CDM}$ model) against $\Omega_m$ and $z$. By visual inspection of Figure \ref{f:funcder} it is clear the relevant impact played by small $z$ and large $\Omega_m$. The functional derivative is hence very small for the portion of the dataset that is populated by the vast majority of SNe entries. This implies that different models (in terms of the associated $I(z)$ ) can yield very similar predictions. It is hence difficult to draw conclusions about the validity of different models, if one solely deals with data at small redshifts.\\

\end{appendix}

\end{document}